\documentclass[longauth]{aa}
\usepackage[varg]{txfonts}
\usepackage{amsmath}
\usepackage{graphicx}
\usepackage{comment}
\usepackage{natbib}

\def\apjl{The Astrophysical Journal, Letters}%
\def\apjs{ApJS}%
\newcommand{\pmi}{$\pm$}
\newcommand{\perc}{$\%$}
\newcommand{\grados}{$^{\circ}$}
\newcommand{\fermi}{\emph{Fermi}-LAT }

\begin{document}

\title{Teraelectronvolt pulsed emission from the Crab pulsar detected
  by MAGIC}

%
\author{
S.~Ansoldi\inst{1} \and
L.~A.~Antonelli\inst{2} \and
P.~Antoranz\inst{3} \and
A.~Babic\inst{4} \and
P.~Bangale\inst{5} \and
U.~Barres de Almeida\inst{5,}\inst{26} \and
J.~A.~Barrio\inst{6} \and
J.~Becerra Gonz\'alez\inst{7,}\inst{27} \and
W.~Bednarek\inst{8} \and
E.~Bernardini\inst{9}
% ...
}
\institute{ Universit\`a di Udine, and INFN Trieste, I-33100 Udine, Italy
\and INAF National Institute for Astrophysics, I-00136 Rome, Italy
\and Universit\`a  di Siena, and INFN Pisa, I-53100 Siena, Italy
\and Croatian MAGIC Consortium, Rudjer Boskovic Institute, University of Rijeka and University of Split, HR-10000 Zagreb, Croatia
\and Max-Planck-Institut f\"ur Physik, D-80805 M\"unchen, Germany
\and Universidad Complutense, E-28040 Madrid, Spain \email dfidalgo@gae.ucm.es
\and Inst. de Astrof\'isica de Canarias, E-38200 La Laguna, Tenerife, Spain
\and University of \L\'od\'z, PL-90236 Lodz, Poland
\and Deutsches Elektronen-Synchrotron (DESY), D-15738 Zeuthen, Germany
\and ETH Zurich, CH-8093 Zurich, Switzerland
\and IFAE, Campus UAB, E-08193 Bellaterra, Spain
\and Universit\"at W\"urzburg, D-97074 W\"urzburg, Germany
\and Centro de Investigaciones Energ\'eticas, Medioambientales y Tecnol\'ogicas, E-28040 Madrid, Spain
\and Institute of Space Sciences (CSIC-IEEC) 08193 Barcelona Spain
\email wilhelmi@aliga.ieec.uab.es 
\and Universit\`a di Padova and INFN, I-35131 Padova, Italy
\and Technische Universit\"at Dortmund, D-44221 Dortmund, Germany
\and Unitat de F\'isica de les Radiacions, Departament de F\'isica, and CERES-IEEC, Universitat Aut\`onoma de Barcelona, E-08193 Bellaterra, Spain
\and Universitat de Barcelona, ICC, IEEC-UB, E-08028 Barcelona, Spain
\email robertazanin@gmail.com; dgalindo@am.ub.es
\and Japanese MAGIC Consortium, Division of Physics and Astronomy, Kyoto University, Japan
\and Academia Sinica, Institute of Astronomy and Astrophysics (ASIAA),
PO Box 23-141, 10617 Taipei, Taiwan,
\and Finnish MAGIC Consortium, Tuorla Observatory, University of Turku and Department of Physics, University of Oulu, Finland
\and Inst. for Nucl. Research and Nucl. Energy, BG-1784 Sofia, Bulgaria
\and Universit\`a di Pisa, and INFN Pisa, I-56126 Pisa, Italy
\and ICREA and Institute of Space Sciences, E-08193 Barcelona, Spain
\and Universit\`a dell'Insubria and INFN Milano Bicocca, Como, I-22100 Como, Italy
\and now at Centro Brasileiro de Pesquisas F\'isicas (CBPF\textbackslash{}MCTI), R. Dr. Xavier Sigaud, 150 - Urca, Rio de Janeiro - RJ, 22290-180, Brazil
\and now at: NASA Goddard Space Flight Center, Greenbelt, MD 20771, USA and Department of Physics and Department of Astronomy, University of Maryland, College Park, MD 20742, USA
% ...
}


\abstract {} 
{To investigate the extension of the very-high-energy
  spectral tail of the Crab pulsar at energies above 400\,GeV.} 
{We analyzed $\sim$320 hours of good quality data of Crab with
  the MAGIC telescope, obtained from February 2007 until April 2014.}
{We report the most energetic pulsed emission ever detected from
  the Crab pulsar reaching up to 1.5\,TeV. The pulse profile shows
  two narrow peaks synchronized with the ones measured in the GeV
  energy range. The spectra of the two peaks
  follow two different power-law functions from 70\,GeV up to 1.5\,TeV
  and connect smoothly with the spectra measured above
  10\,GeV by the Large Area Telescope (LAT) on board of the
  \emph{Fermi} satellite. When making a joint fit of the LAT and MAGIC
  data, above 10\,GeV, the photon indices of the spectra differ by 0.5$\pm$0.1.}
{We measured with the MAGIC telescopes the most energetic pulsed
  photons from a pulsar to date. Such TeV pulsed photons require a
  parent population of electrons with a Lorentz factor of at least
  $5\times 10^6$. These results strongly suggest IC scattering off low energy photons as the emission mechanism and a gamma-ray
  production region in the vicinity of the light cylinder.}

		
\maketitle

\section{Introduction}

The Crab pulsar, PSR J0534+220, is a young neutron star (NS) with a
rotational period of 33\,ms. It was created after the supernova explosion
SN1054.
The Crab is the most powerful pulsar in our Galaxy, with a spin-down
luminosity of $4.6 \times 10^{38}$ erg\,s$^{-1}$. It is one of the few pulsars
detected across the electromagnetic spectrum, from radio up to gamma
rays, and one of the brightest at high energies (HE ,
0.1$<$E$<$10\,GeV, \citealt{fierro:1998:egret:crabp,
  kuiper:2001:comptel:crabp,abdo:2010:fermi:crab,aliu:2008:magic:crabp}).
The exceptionality of this source was recently underlined by the
discovery of pulsed emission at energies up to 400\,GeV
\citep{aliu:2011:veritas:science,aleksic:2012:magic:pulsar400}.

The Crab pulsar emission profile is characterized by three components:
two pulses separated by $\sim$0.4 in phase observed at all energies; from
centimeter radio (E$\sim$10$^{-4}$\,eV) to very-high-energy gamma rays (VHE, E$>$\,100\,GeV), and a third component, the bridge, which is defined
as the pulse phase between the main pulse (P1) and the second pulse
(P2). P1 has the highest intensity at radio frequencies and defines
phase 0, whereas P2, which is often referred to as the interpulse,
is weaker at radio frequencies. The amplitude of each pulse depends on
the energy \citep{kuiper:2001:comptel:crabp}, and in particular, in
the gamma-ray regime, P2 becomes dominant above 25-50\,GeV,
whereas the bridge is only detected up to 150\,GeV
\citep{aleksic:2014:magic:bridge}.

The HE gamma-ray emission from
pulsars is believed to be produced via synchrotron-curvature radiation
by electron-positron pairs moving along curved paths inside the
light cylinder. The maximum photon energy is
limited by either magnetic and gamma-gamma pair absorption or radiation losses,
resulting in spectral cutoffs at around a few GeV
\citep{cheng:1986:outergap}. This theoretical scenario has
been confirmed by the analysis of about 150 pulsars detected by the
\fermi gamma-ray telescope \citep{abdo:2013:pulsarcatalogue}.  The observed pulse
profiles and spectral shapes suggest that the gamma-ray beams have a
fan-like geometry and that they are located at high-altitude zones
inside the magnetosphere, towards the spin equator: either close to
the light cylinder (LC, outer gap models,
\citealt{cheng:1986:outergap,romani:1995:outergap,cheng:2000:outergap,takata:2006:outergap})
or along the last open magnetic field lines (slot gap models,
\citealt{arons:1983:slotgap,muslimov:2004:slotgap}).

The first year of \fermi observations of the Crab pulsar spectrum
validates the consensus view reporting a spectral cutoff at (5.8\pmi
0.5$_{stat}$\pmi 1.2$_{syst}$)\,GeV
\citep{abdo:2010:fermi:crab}. However, the gamma-ray emission later
discovered at VHE \citep{aliu:2011:veritas:science,aleksic:2011:magicST:crab,aleksic:2012:magic:pulsar400},
is not compatible (at more than a 6$\sigma$ confidence level) with flux
predictions based on synchro-curvature emission. 
This new and unexpected spectral component, described by a steep
power-law function (with a photon index of approximately 3.5) between
25 and 400\,GeV, required an ad-hoc explanation
\citep{aliu:2011:veritas:science,aleksic:2011:magicST:crab,aleksic:2012:magic:pulsar400}.
Some of the advocated models include the same synchro-curvature
mechanism responsible for the sub-TeV emission, yet under extreme
conditions \citep{bednarek:2012:crabp,vigano:2015:crabp}, whereas others proposed a
new mechanism to be at work: inverse Compton (IC) scattering on seed photon
fields (from infrared to X-rays). In the case of IC
radiation, different VHE gamma-ray production regions have been
considered; from the acceleration gap in the pulsar magnetosphere
\citep{aleksic:2011:magicST:crab,hirotani:2011:crabp,lyutikov:2012:crabp,2015arXiv150806251H}
to the ultra-relativistic cold wind that extends from the light cylinder to
the wind shock
\citep{aharonian:2012:crabp,bogovalov:2000:crabp,mochol:2015:crabp:stripedwind}.

The goal of this work is to investigate the maximum energy reached in
the Crab pulsar spectrum. For this purpose, we reanalyzed more than 300 hours of excellent quality data of the Crab recorded by MAGIC from 2007 to 2014, both in stand-alone and stereoscopic mode.

\begin{figure}[!th]
\centering
\includegraphics[width=0.4\textwidth,height=5cm]{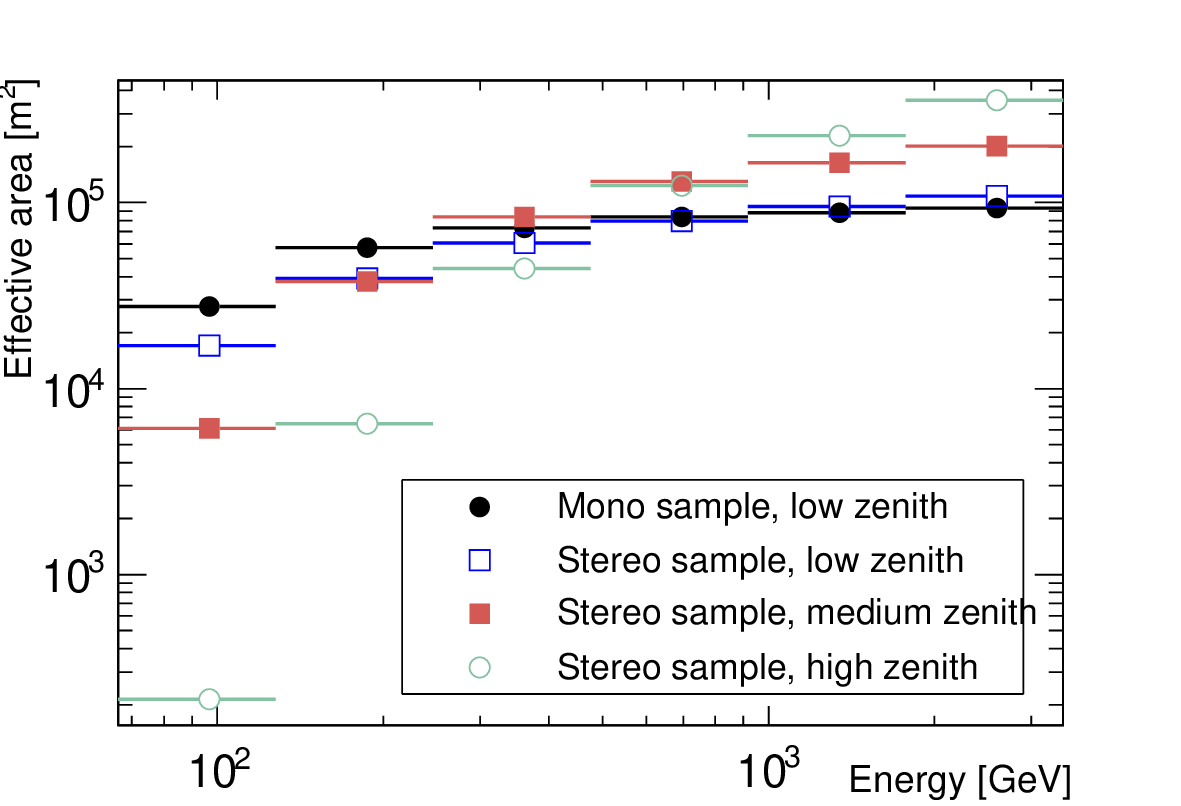}
\caption{Effective area, after background rejection cuts, for four
  representative data subsamples.\label{fig:area}}
\end{figure}

\section{Observations and analysis}
MAGIC is an array of two imaging atmospheric Cherenkov telescopes
(IACTs) designed for the detection of gamma rays in the energy band
between few tens of GeV and few tens of TeV. 
It is located in the Canary island of La Palma (Spain) at 2250\,m above
the sea level. 
The two telescopes have a 17\,m diameter reflective surface and fine
pixelized cameras with a 3.5\grados{} field of view. 

MAGIC started its operations in stand-alone mode with the first MAGIC
telescope, MAGIC-I, in autumn 2004. 
In February 2007, its readout system was upgraded to an ultrafast
FADC of 2 GHz sampling rate \citep{goebel:2008:mux} which allowed for 
a better reconstruction of the timing characteristics of the recorded
images and for a factor of two background reduction. 
The upgraded MAGIC-I could detect sources with fluxes as low as
1.6\perc{} of the Crab nebula flux above 280\,GeV in 50 hours of
observation \citep{aliu:2009:magic:timing}. It had an energy
resolution of 20\perc{} at around 1\,TeV. Observations carried out during
this initial phase will be referred to as mono observations hereafter.
In 2009, MAGIC became a stereoscopic system leading
to an improvement in sensitivity by a factor 2
\citep{aleksic:2012:magic:stereo1:performance}. To equalize the
performance and hardware of the two telescopes, a major upgrade was carried out
during the summers of 2011 and 2012. First, the readout systems
of both telescopes were upgraded with the domino ring sampler version
4 chip, and, in the following year, the MAGIC-I camera was
replaced by a uniformly pixelized one, mainly a clone of the second
telescope camera \citep{aleksic:2015:magic:majorupgrade:hardware}. Currently
the array has an energy threshold as low as $\sim$70\,GeV, for low zenith angle
observations, and an integral sensitivity above 300\,GeV of 0.6\perc{}
of the Crab nebula flux in 50 hours of observation
\citep{aleksic:2015:magic:majorupgrade:performance}. The energy resolution
is 15--17\perc{} at $\sim$1\,TeV. 

The analysis was performed by using the standard MAGIC
software, MARS \citep{zanin:2013:magic:mars}. The gamma/hadron
separation and the estimation of the gamma-ray direction make use of
random forest (RF) algorithms \citep{albert:2008:magic:RF,aleksic:2010:magic:agnhalo}. 
The energy estimation can be performed either by means of the RF technique
or with Monte Carlo (MC) look-up tables (LUTs), which are the standard
procedures for mono and stereo data analysis, respectively.
In the case of the Crab pulsar above $\sim$100\,GeV the background is no
longer dominated by hadrons, but gamma rays from the Crab
nebula. Therefore, we applied background rejection cuts specifically
optimized for a gamma-dominated background and specified that at least
90\perc{} of our MC gamma rays survive those cuts. The cut
optimization is based on the maximization of the modified formula (17) by \citealt{lima:1983}
which considers as background the hadronic and nebula events derived
from the nebula excess and the power-law spectrum for the pulsar found in \citealt{aleksic:2012:magic:pulsar400}.
For the differential energy spectra, we applied an unfolding procedure
correcting for the energy bias and the detector finite energy
resolution. We tested the five unfolding methods described in 
\citet{albert:2007:magic:spectrumunfolding} and verified their
consistency within statistical errors. 
The upper limits (ULs) to the differential flux were obtained by
following the \citet{rolke:2005} method under the assumption of a Gaussian
background and 20\perc{} systematic uncertainty in the flux
level. Hereafter, the ULs will be given at 95\perc{} confidence level (CL). 
The pulsar rotational phase of each event was defined by using 
the TEMPO2 package \citep{hobbs:2006:tempo2} (and cross-checked by our
own code, \citealt{lopez:2006:psearch}) and the
monthly ephemeris publicly provided by the Jodrell Bank
Observatory\footnote{http://www.jb.man.ac.uk/research/pulsar/crab.html}
\citep{lyne:1993:jodrell}.

In this work we used all the data taken in stereoscopic mode, until April
2014, when pointing at the Crab. The selected sample includes
observations performed at zenith angles up to 70\grados{}. In order to
increase the statistics we also reanalyzed Crab mono data recorded 
after the upgrade of the readout system (between 2007 and April 2009)
at zenith angles smaller than 30\grados{}.
Both mono and stereo data samples were taken partially in ``on'' and,
partially, in false-source tracking \citep{fomin:1994:wobble} mode, the
latter pointing at two symmetric positions 0.4\grados{} away from the source. 
Data affected by hardware problems, bad atmospheric
conditions, and showing unusual hadron rates were removed from the
analyzed data sample, resulting in 97 hours and 221 hours of effective
time for the mono and the stereo samples, respectively. 
Given that the considered data sample spreads over seven years, with
different instrument performance, we divided it into nine analysis
periods, each with its corresponding MC simulation. The whole
data sample was then further subdivided into three zenith angle ranges
to better account for the corresponding dependence of the image shower
parameters at the cut optimization stage. This resulted in 19 data
sub-samples, each period with at least some low zenith angle data used to monitor the instrument performance. RF matrices and energy LUTs were
produced separately. Figure\,\ref{fig:area} shows the effective area
for four representative datasets: mono, and stereo in the three zenith
angle ranges. The differential energy spectra obtained for each
independent analysis were later on combined, once weighted with the
exposure, when applying the unfolding procedure.

\section{Results}

\subsection{Light curve}

\begin{figure}[!th]
\centering
\includegraphics[width=0.5\textwidth,height=9cm]{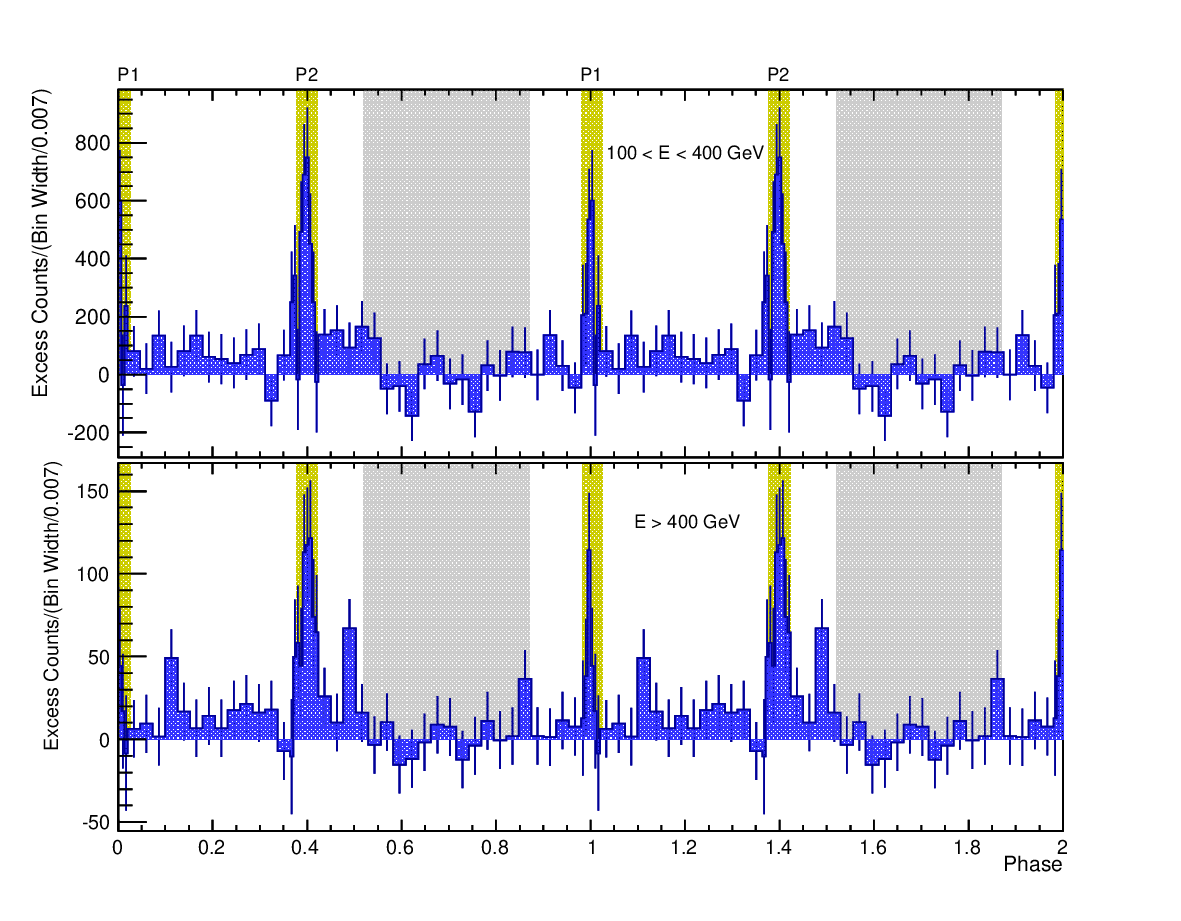}
\caption{Pulse profile of the Crab pulsar
  between 100 and 400\,GeV (upper panel) and above 400\,GeV (bottom
  panel). The pulse profile, shown twice for clarity, is background
  subtracted. The bin width around the two peaks is 4 times smaller
  (0.007) than the rest (0.027) in order to highlight the sharpness of
  the peaks. Yellow-dashed areas identify the phase intervals of the
  two peaks, whereas the gray areas show the off-pulse region.\label{fig:lc}}
\end{figure}

In the search for pulsation above 400\,GeV from the Crab pulsar we
defined the phase ranges of the two peaks according to the results
obtained in our previous studies
\citep{aleksic:2012:magic:pulsar400,aleksic:2014:magic:bridge}: the
main peak P1 $\in$ ($-$0.017 -- 0.026) and the interpulse P2 $\in$ (0.377 --
0.422). The interval (0.52 -- 0.87) was considered as off-pulse
region \citep{fierro:1998:egret:crabp} from where we
estimated the background to be subtracted from the histograms.

Figure \ref{fig:lc} shows the folded pulse profile that we obtained
between 100 and 400\, GeV and above 400\,GeV with 318 hours of
observation. 
In the 100--400\,GeV energy range P1 is detected with a significance level of 
2.8$\sigma$, whereas P2 at 5.6$\sigma$ after \citet[Eq.17]{lima:1983}. The
statistical significance of the detection of P1 and P2 with this
analysis is smaller than that reported in
\citet{aleksic:2014:magic:bridge} with less than half of the
observation time. This is due to the fact that the analysis presented
in this work combines many periods with different sensitivities and energy thresholds, and these factors contribute to decreasing the signal-to-noise ratio at the lowest energies, hence
worsening the overall signal significance. If we consider only stereo data
for zenith angles below 35\grados{}, which identify the data sub-sample
with the lowest energy threshold and best gamma/hadron separation at
the lowest energies, we end up with 152 hours of observation time,
yielding a signal significance of 6.6$\sigma$ and 8.8$\sigma$ for P1
and P2 respectively in the energy range between 100 and 400\,GeV. This
is in agreement with the results reported in
\citet{aleksic:2014:magic:bridge} for the 50--400\,GeV energy range.
Beyond 400\,GeV (above the energy threshold of all the 19
analyses used here) the gamma/hadron separation is efficient for all
the analyses and we have a clear gain in the signal significance for
the combined sample due to the increase in photon statistics. 
For energies above 400\,GeV, only P2 is significantly detected. 
The total number of excess events are 544\pmi92 and 188\pmi 88 for P2 and P1 respectively corresponding to 6$\sigma$ and 2.2$\sigma$ for each peak.
With a higher energy cut at 500\,GeV, meant
to exclude the lower energy events from the light curve where no
spillover correction is applied, P2 is still detected at 5$\sigma$,
while P1 shows a 2$\sigma$ signal with 418\pmi104 and 152\pmi108
excess events, respectively. Table\,\ref{tab:sigma} summarizes the
number of excess events with their corresponding significance for
different integral energy ranges.\\
The significance of the pulsation was also tested with the H-test
\citep{dejager:1989:Htest} which does not make any a priori assumption on
the position and the shape of the pulsed emission, resulting in a
3.5 (2.8)$\sigma$ significance above 400 (500)\,GeV.

\begin{table}[h]
\centering
\footnotesize
\caption{Number of excess events and corresponding significance of P1
  and P2 for different energy ranges in $\sim$320 hours of data.
\label{tab:sigma}}
\begin{tabular}{c|c|c|c|c}
\hline
\hline
energy range  & \multicolumn{2}{c|}{P1} & \multicolumn{2}{c}{P2}\\
\hline
[GeV] & N$_{\rm{ex}}$ & Significance &  N$_{\rm{ex}}$ & Significance \\
\hline
100--400 & 1252$\pm$442 & 2.8 $\sigma$ & 2537$\pm$454 & 5.6 $\sigma$\\
$>$ 400 & 188$\pm$88 & 2.2 $\sigma$ & 544 $\pm$ 92 & 6.0 $\sigma$\\
$>$ 680 &  130$\pm$66 & 2.0 $\sigma$ & 293$\pm$69 & 4.3 $\sigma$\\
$>$ 950 & 119$\pm$54 &2.2 $\sigma$  & 190$\pm$56 & 3.5 $\sigma$\\
\hline
\hline
\end{tabular}
\end{table}


We fitted the pulse profile above 400\,GeV to a finer-binned
distribution with two symmetric Gaussian functions (as in
\citealt{aleksic:2012:magic:pulsar400}). The available statistics does
not allow us to consider more complicated functions. {P1 and P2 are
located at the phases 0.9968$\pm$0.0020$_{stat}$+0.0055$_{syst}$$-$0.0048$_{syst}$ and
0.4046$\pm$0.0035$_{stat}$+0.0047$_{syst}$$-$0.0074$_{syst}$ respectively, in agreement
with the positions found at lower energies between 50 and 400\,GeV
\citep{aleksic:2012:magic:pulsar400}. The full width at half maximum
(FWHM) for P1 is 0.010$\pm$0.003$_{stat}$+0.003$_{syst}$$-$0.010$_{syst}$ and for
P2 is  0.040$\pm$0.009$_{stat}$ + 0.005$_{syst}$$-$0.008$_{syst}$. The systematic
uncertainty on the estimation of the peak positions reflects the
precision of the pulsar ephemerides used for this analysis, taking
into account the RMS of the timing noise, the uncertainty on the
arrival time of the first pulse taken as reference, and the error
introduced by the barycentric corrections. It also includes the effect of the histogram binning. The width of the peaks beyond this energy is compatible within the errors with the value measured below 400 GeV.  Note that results reported above 400\,GeV for P1 are
obtained for a $\sim$2$\sigma$ signal and should be taken with caution. For comparison, the best-fit P1 and P2
positions in the 100\,MeV to 10\,GeV energy range are 0.9915$\pm$0.0005
and 0.3894$\pm$0.0022 \citep{abdo:2010:fermi:crab}. 

\subsection{Energy spectra}

Figure \ref{fig:sed} shows the phase-folded spectral energy
distributions (SED) of P1 and P2 from $\sim$70\,GeV up to 1.5\,TeV,
obtained by using the Bertero's unfolding method \citep{bertero:1989:unfolding}. 
Both the differential energy spectra are well-described by power-law
functions with a photon index, $\alpha$, of
$3.2\pm0.4_{stat}\pm0.3_{syst}$ and
$2.9\pm0.2_{stat}\pm0.3_{syst}$, for P1 and P2,
respectively. The results of the fits, shown in Table\,\ref{tab:fit}
for a normalization energy at 150\,GeV (being the decorrelation energy 120 GeV and 190 GeV for P1 and P2 respectively),
are in agreement with our earlier results
\citep{aleksic:2012:magic:pulsar400, aleksic:2014:magic:bridge}.
In the case of P2, the power-law spectrum extends up to 1.5\,TeV,
whereas P1 cannot be measured beyond 600\,GeV.
At energies above the last obtained spectral point, we computed ULs to
the differential flux, at 95\% CL, under the assumption of the
power-law spectrum found in this work. However, a 20\perc{} change in the
photon index yields a variation of less than 15\perc{} in the UL.  These
ULs are not constraining any possible cutoff, given the current
sensitivity of the instrument. The spectral points and ULs are listed
in Table\,\ref{tab:points}. 

\begin{figure}[!t]
\centering
\includegraphics[width=0.5\textwidth]{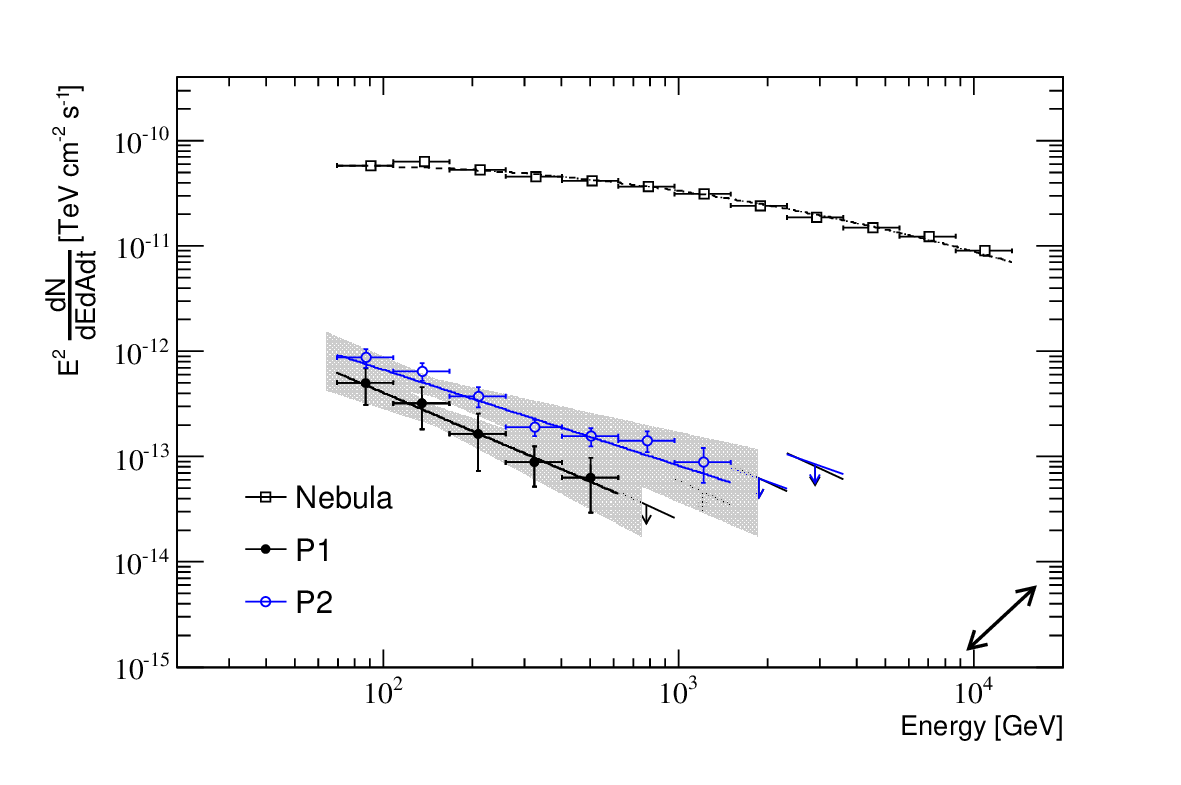}
\caption{Phase-folded SED of the Crab P1 (black circles) and P2 (blue circles) measured by MAGIC between $\sim$70\,GeV and 1.5\,TeV. The
  butterfly identifies the systematic uncertainty on the flux
  normalization and spectral index, whereas the arrow on the bottom
  right corner corresponds to an energy shift of 17\perc{}. The Crab
  nebula spectrum (open squares) is also shown for comparison. The
  differential flux upper limits, at  95\perc{} CL, are computed under
  the assumption of the power-law spectrum measured in this work.
\label{fig:sed}}
\end{figure}

\begin{figure}[!t]
\centering
\includegraphics[width=0.5\textwidth]{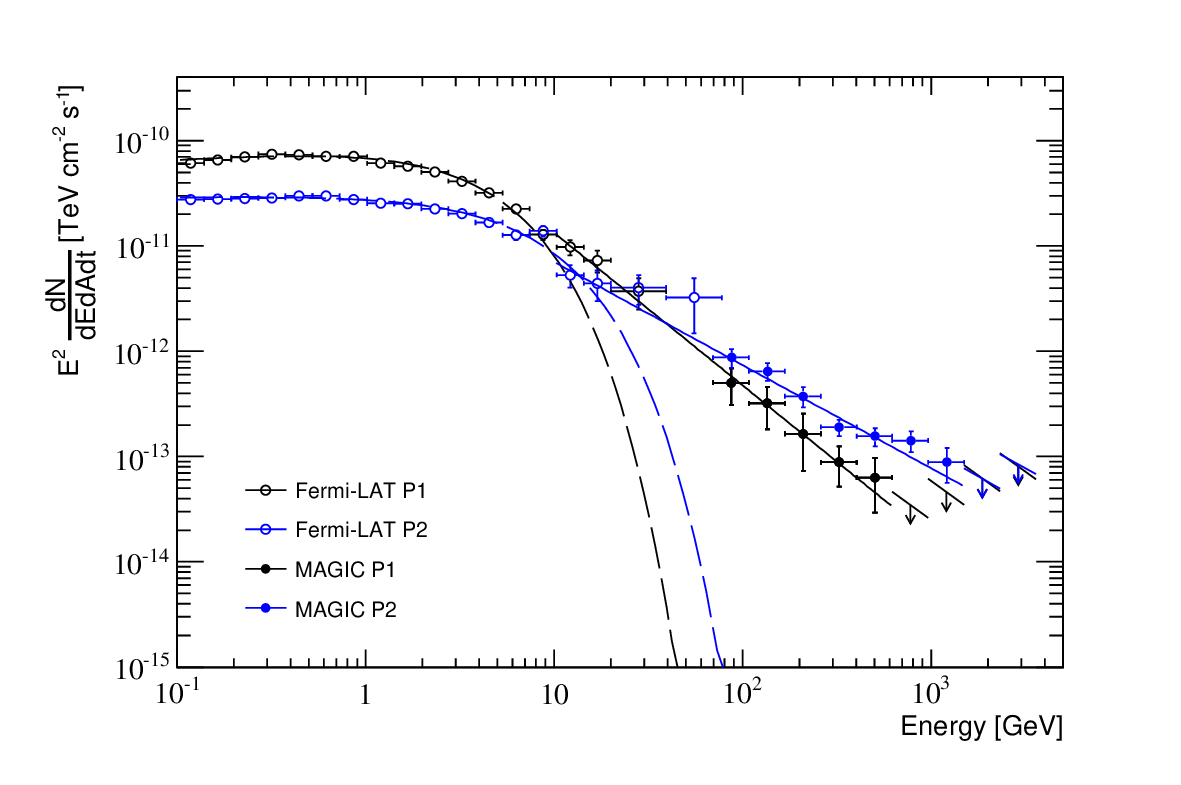}
\caption{Phase-folded SED of the Crab P1 (black circles) and P2 (blue
  circles) at HE and VHE (open and filled circles). The results of the
  power-law with exponential cutoff fits to the \fermi points are
  illustrated by the dashed lines \citep{aleksic:2014:magic:bridge},
  whereas the joint \emph{Fermi}-LAT/MAGIC fits
  to power-law functions above 10\,GeV are shown by solid lines. The
  upper limits to the differential flux, at 95\perc{} CL, are computed
  under the assumption of the power-law spectrum found in this work,
  as represented by the slope of the arrows. \label{fig:sedfermi}}
\end{figure}

The extrapolation of the MAGIC energy spectra to lower energies agrees
within the statistical errors with the spectra measured with \fermi
above 10\,GeV. The latter was already showing a
deviation from the expected exponential cutoff
\citep{aleksic:2014:magic:bridge}. A joint
correlated-$\chi^2$-fit\footnote{The fit takes into account the
  correlation between the MAGIC spectral points due to the unfolding
  procedure.} of MAGIC and \fermi spectral points above 10\,GeV shows
that the new spectral components are well-represented ($\chi^2/ndf$=1.5/6 and
$\chi^2/ndf$=8.5/9 for P1 and P2, respectively) by simple
power-law functions (see Table\,\ref{tab:fit}), where the normalization
energy is set to 50\,GeV. The photon
indices of the two power-law functions are $\alpha=3.5\pm0.1$ and
$\alpha=3.0\pm0.1$ for P1 and P2,
respectively. The difference in the spectral slopes by $\Delta
\alpha=0.5\pm0.1$ is significant by more than 3$\sigma$, indicating
that the intensity of P1 drops more rapidly with energy than that of
P2. At X-ray energies (3--10\,keV) \emph{NuSTAR} detected a similar
spectral behaviour with P2 harder than P1 and the corresponding photon
indices being 1.66$\pm$0.02 and 1.80$\pm$0.01, respectively
\citep{madsen:2015:crabp:nustar}. A fit to a power-law function plus exponential cutoff allows us to
impose a lower limit in the spectral cutoff of 700\,GeV at 95\% CL.

The measured spectral difference at VHE could be naturally explained either by two distinct
production locations for each peak or by the difference in the
phase-resolved spectrum of X-rays which act as targets for IC scattering.

We cross-checked the P2 energy spectrum by comparing mono data
to the stereo data and found that the results were stable within
statistical errors for all the considered unfolding methods.
We also computed the Crab nebula SED, as shown in Figure \ref{fig:sed}
(open squares),
using the subsample of the data taken in wobble mode. The nebula
spectral measurement was obtained by analyzing the same energy range as the pulsar analysis, using the same energy binning and gamma selection cuts. The resulting
spectral points are consistent with the results presented in \citet{aleksic:2012:magic:stereo1:performance,aleksic:2015:magic:mio}. 
Therefore, we assumed that no extra systematic uncertainty on the
total flux is needed for this specific analysis. These systematic
uncertainties are 17\perc{} on the energy scale, 19\perc{} on the flux
normalization, and 0.3 on the photon index. The latter error is the
only one not in agreement with
\citet{aleksic:2012:magic:stereo1:performance}, and mainly arises 
from the larger uncertainty of the unfolding given the low
statistics of the result.

\begin{table}[h]
\centering
\footnotesize
\caption{Results of the spectral fit to a power-law function. Errors
  indicate 1$\sigma$ statistical uncertainties. E$_{\rm{o}}$
  indicates the normalization energy.}
\label{tab:fit}
\begin{tabular}{c|c|c|c|c|c}
\hline
\hline
&  & E$_{\rm{o}}$ & $f_{\rm{E_{o}}}$  & $\alpha$ & $\chi^2$/dof\\
& & [GeV] & [TeV$^{-1}$ cm$^{-2}$ s$^{-1}$] & & \\
\hline
MAGIC & P1 & 150  & (1.1\pmi0.3)$\times$10$^{-11}$&3.2\pmi0.4 & 0.3/3 \\
&  P2 & 150 &  (2.0\pmi0.3) $\times$10$^{-11}$ &2.9\pmi0.2&5.4/5\\ 
\hline 
\fermi{} & P1 & 50  & (5.3\pmi0.8)$\times$10$^{-10}$&3.5\pmi0.1 & 1.5/6
\\
\& MAGIC & P2 & 50 & (5.7\pmi0.6) $\times$10$^{-10}$ &3.0\pmi0.1&8.4/9\\ 
\hline
\hline
\end{tabular}
\end{table}

\begin{small}
\begin{table}
\centering  
\footnotesize
\caption{Spectral points of the MAGIC measurements shown in Figure \ref{fig:sed}. \label{tab:points}  }
\begin{tabular}{c|c|c|c}
\hline
\hline
& & P1 & P2 \\
Energy & Bin Center & E$^{2}$dN/dEdAdt &  E$^{2}$dN/dEdAdt \\
$[\rm{GeV}]$ & [$\rm{GeV}$] & [TeV\,cm$^{-2}$\,s$^{-1}$] &
[TeV\,cm$^{-2}$\,s$^{-1}$] \\
& & $\times 10^{-13}$ & $\times 10^{-13}$\\
\hline
69--108 & 87 & 5.0$\pm$ 1.9& 8.7 $\pm$ 1.8 \\
108--167 & 135 &3.2 $\pm$ 1.4 & 6.5 $\pm$ 1.2 \\
167--259 & 210 &1.6 $\pm$ 0.9 &3.7 $\pm$ 0.8 \\
259--402 & 325 &0.9$\pm$0.4  & 1.9$\pm$0.3 \\
402--623 & 504 & 0.6$\pm$0.3 & 1.6 $\pm$ 0.3 \\
623--965 & 781 & $<$ 0.3 & 1.4$\pm$ 0.3\\
965--1497 & 1211 & $<$0.5 & 0.9 $\pm$0.3\\
1497--2321 & 1879 & $<$0.6 & $<0.6$ \\
2321-3598 & 2914 & $<$0.8 & $<0.8$ \\
\hline
\hline
\end{tabular} 
\end{table}
\end{small}
\section{Discussion and conclusions}

The new results presented here probe the Crab pulsar as the most
compact TeV accelerator known to date.
The remarkable detection of pulsed emission up to 1.5\,TeV revealed
by MAGIC imposes severe constraints on where and how the underlying
electron population produces gamma rays at these energies. The electron population responsible for the VHE emission should have Lorentz factors greater than $5\times10^6$, which can be responsible for the VHE emission only when accelerated near or beyond the light cylinder \citep{bogovalov:2014}. The TeV pulsed emission cannot be produced with synchro-curvature radiation, even in the extreme case in which the
magnetic-field-aligned electric field approaches the strength of the
magnetic field. In this scenario, the curvature radius would have to
be one order of magnitude larger than the typical one, which is
believed to be between 0.3 and 2 times the light cylinder radius
\citep{vigano:2015:outergap} (for further discussions on this issue we refer to \citealt{2014ApJ...793...97K,2015arXiv150806251H} and references therein). 
Therefore, the unprecedented measurement of pulsed emission extending
up to TeV energies performed with the MAGIC telescopes implies that
the IC process is at work in the Crab pulsar, and that
it dominates the emission of gamma rays above 50\,GeV. This 
partially solves the puzzle posed by the previous published results but also
opens new challenges. We note that although other processes
(e.g. synchro-curvature radiation) could account for the production of
100--400\,GeV photons, the simple power-law function obtained by a
joint fit of \fermi and MAGIC data from $\sim$10\,GeV up to 1.5\,TeV suggests
a single mechanism for both P1 and P2,  and that this must
be Compton up-scattering of soft photons off high-energy electrons.

Concerning IC scattering, two scenarios which were previously
proposed to explain the VHE emission below 400\,GeV can
be considered: the magnetospheric synchrotron-self-Compton model
\citep{aleksic:2011:magicST:crab} and the IC in the pulsar wind region model
\citep{aharonian:2012:crabp}.
The former assumes that there are acceleration gaps in the outer
magnetosphere
\citep{cheng:1986:outergap,romani:1995:outergap,cheng:2000:outergap,takata:2006:outergap,arons:1983:slotgap,muslimov:2004:slotgap}
where primary  positrons propagate outwards and escape, being illuminated by a strong
magnetospheric infrared (IR) photon field which is then up-scattered by 
positrons to TeV-scale energies. These primary TeV photons are then
efficiently absorbed by the same IR field to materialize as secondary
e$^{\pm}$ pairs with GeV to several TeV energies. Such secondary pairs
are created at a greater distance whereby there is a lower photon-field
density, near to and outside the LC, and can up-scatter the IR-UV
photons into 10\,GeV--5\,TeV photons (via synchrotron self-Compton process,
\citealt{hirotani:2013}). Some of them escape from the magnetosphere and are
observable from Earth. However the synchronization of the pulse profile in the GeV and TeV
regimes limits this interpretation, suggesting a similar region of
generation, where absorption of TeV photons is unavoidable. The
measured time delay between the best-fit peak positions in the MeV--GeV and
the TeV regime is 178$\pm$69$\mu$s and 512$\pm$139$\mu$s for P1 and P2
respectively, which, when considering the relatively large systematics in the determination of the peak positions, are compatible with the hypothesis of no separation between the bulk of the radiation region where all these photons are generated (neglecting more complicated geometrical effects and assuming the simple case of stationary emission regions).

The pulsar wind scenario considers the IC scattering off the
synchrotron, pulsed IR and X-ray photons by the particles
(electron/positron) of the cold relativistic wind.
It is commonly accepted that the pulsar wind is magnetically dominated
near the LC. Thus, in the wind model, the wind becomes abruptly
particle-kinetic-energy dominated over a short distance
(compared to the dimension of the wind region). Based on previous
results by Cherenkov telescopes on the Crab pulsar
\citep{aliu:2011:veritas:science,aleksic:2012:magic:pulsar400}, this
distance was estimated to be 20-50 LC radii
\citep{aharonian:2012:crabp}. In this narrow cylindrical zone, electrons and positrons are rapidly
accelerated up to Lorentz factors of $5\times 10^5$. 
The bulk Lorentz factor is assumed to display a power-law dependence
on the distance: $\Gamma(R)$=$\Gamma_0 +
(\Gamma_w -\Gamma_0)(\frac{(R-R_0)}{(R_{f}-R_0)})^{\alpha}$, 
where $\Gamma_0$ and $\Gamma_w$ are the initial and the maximum wind
Lorentz factors, $R_0$ the distance at which the acceleration starts,
$R_f$ the distance at which $\Gamma_w$ is reached, and the power-law
index $\alpha \sim $1,3,10 \citep{aharonian:2012:crabp}. 
To obtain a $\Gamma_w$ compatible with the one that is derived from
the detection of TeV photons ($\sim\,5 \times 10^6)$, the region in
which particles are accelerated has to extend up to a much larger
radius than the one considered in \citealt{aharonian:2012:crabp}. In
this case however, the model fails to reproduce the spectral shape
below 100\,GeV \citep[Figure SM1]{aharonian:2012:crabp}. Instead, a slower and
continuous acceleration (for instance due to magnetic 
reconnection) or a more complex radial dependence could be at
play. Other approaches in the context of the pulsar wind emission
region and/or pulsar magnetosphere are currently being investigated to try to
give a satisfactory explanation to the  TeV pulsed emission
\citep{mochol:2015:crabp:stripedwind,2015arXiv150806251H}. 
So far all the existing models fail to reproduce the narrow peaks
\citep{aleksic:2011:magicST:crab,aharonian:2012:crabp} observed in the
Crab pulsar light curve above 400\,GeV, keeping the coherence along
four decades in energy.
The MAGIC results require a revision of the state-of-the-art models
proposed to explain how and where gamma ray pulsed emission from
100\,MeV to 1.5\,TeV are produced.

\begin{acknowledgements} 
We would like to thank
the Instituto de Astrof\'{\i}sica de Canarias
for the excellent working conditions
at the Observatorio del Roque de los Muchachos in La Palma.
The financial support of the German BMBF and MPG,
the Italian INFN and INAF,
the Swiss National Fund SNF,
the ERDF under the Spanish MINECO (FPA2012-39502), and
the Japanese JSPS and MEXT
is gratefully acknowledged.
This work was also supported
by the Centro de Excelencia Severo Ochoa SEV-2012-0234, CPAN CSD2007-00042, and MultiDark CSD2009-00064 projects of the Spanish Consolider-Ingenio 2010 programme,
by grant 268740 of the Academy of Finland,
by the Croatian Science Foundation (HrZZ) Project 09/176 and the University of Rijeka Project 13.12.1.3.02,
by the DFG Collaborative Research Centers SFB823/C4 and SFB876/C3,
and by the Polish MNiSzW grant 745/N-HESS-MAGIC/2010/0.
\end{acknowledgements}

\bibliographystyle{aa}
\bibliography{crabVHE}


\end{document}